\documentclass[11pt]{article}
\usepackage{moriond,epsfig}

\def\be{\begin{equation}}
\def\ee{\end{equation}}
\def\bea{\begin{eqnarray}}
\def\eea{\end{eqnarray}}

\begin{document}
\vspace*{4cm}
\title{\boldmath$J/\psi$ PRODUCTION IN DIRECT AND RESOLVED PHOTON COLLISIONS}

\author{ M. KLASEN }

\address{II. Institut f\"ur Theoretische Physik,
        Universit\"at Hamburg, \\
        Luruper Chaussee 149, D-22761 Hamburg, Germany}

\maketitle\abstracts{
The production of $J/\psi$ mesons in photon photon collisions allows for a
test of factorization in color-octet processes as predicted by NRQCD and
observed at the Tevatron. We calculate the cross sections for $J/\psi$
production with direct and resolved photons, including also the feed-down from
$\chi_{cJ}$ and $\psi^\prime$ decays. Our NRQCD predictions are nicely
confirmed by recent data from the DELPHI collaboration at CERN LEP2.
}

\section{Motivation}
One of the most interesting features of quantum chromodynamics (QCD) is the
apparent discrepancy between the fundamental role of color as a conserved
quantum number carried by quarks and gluons and the non-observation of color
in physical hadronic states. As a non-perturbative phenomenon, the confinement
mechanism responsible for the color-singlet nature of hadrons is still not
fully understood. One can hope to gain more insight by studying the relatively
simple bound states of heavy charm and bottom quarks, such as $J/\psi$ or
$\Upsilon$ mesons.

Ever since QCD emerged as the fundamental theory of the strong interaction,
models have been designed to explain the compensation of color in heavy
quarkonia. Well-known examples are the Color Evaporation
\cite{Fritzsch:1977ay,Halzen:1977rs,Gluck:1978bf} and Hard Comover Scattering
Models \cite{Hoyer:1998ha,Marchal:2000wd}, which assume that color is restored
by soft and hard gluon exchange with the underlying event, respectively, and
the Color Singlet Model (CSM) \cite{Berger:1980ni,Baier:1981uk}, which requires
the heavy quark pair to be produced in a color singlet state. The CSM leads to
strong
selection rules and connects the production cross section with the quarkonium
wave function. Today, a rigorous effective field theory exists in the form of
non-relativistic QCD (NRQCD) \cite{Caswell:1985ui}, which utilizes a double
expansion in the strong coupling constant $\alpha_s$ and the relative
quark-antiquark velocity $v$ in order to factorize the hard production from
the soft binding process \cite{Bodwin:1994jh}. Left-over singularities in the
CSM can be removed systematically into non-perturbative
color-octet operator matrix elements (OMEs), and their numerical values can be
fitted to describe the hadroproduction cross sections observed at the Tevatron.
However, it is necessary to demonstrate the universality of these OMEs in
other production processes, such as $\gamma\gamma$ collisions.
Since the theoretical uncertainties from scale variations in leading order
(LO) of $\alpha_s$ are quite substantial, it is furthermore important to
include also
contributions from next-to-leading order (NLO) virtual loop and real emission
processes. In this paper, we report on recent progress along these lines.

\section{Real corrections to $J/\psi$ production in direct $\gamma\gamma$ collisions}

\begin{figure*}
\begin{center}
\epsfig{file=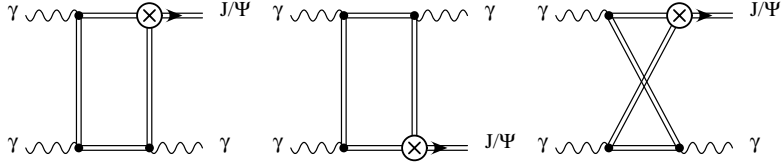,bbllx=60pt,bblly=360pt,bburx=380pt,bbury=431pt,%
        width=0.65\textwidth}
\end{center}
\caption{\label{fig:1}Feynman diagrams pertinent to the partonic subprocess
         $\gamma\gamma\to J/\psi\gamma$.}
\end{figure*}

\begin{figure*}[htb]
\begin{center}
\epsfig{file=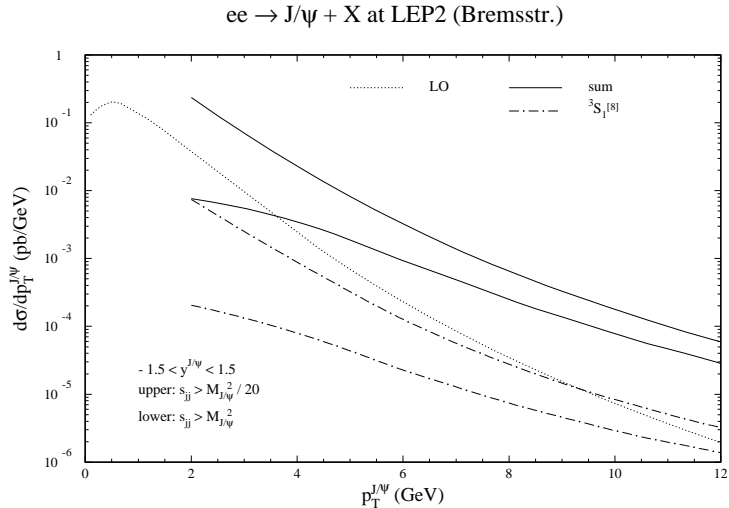,width=0.65\textwidth}
\end{center}
\vspace*{-10mm}
\caption{\label{fig:2}Transverse-momentum distribution $d\sigma/dp_T$ of
$\gamma\gamma\to J/\psi+X$ via bremsstrahlung at CERN LEP2.
The sum of the LO contributions for $X=j$ and $X=\gamma$ is compared with the
$2\to3$ part of the NLO contribution for dijet mass $s_{jj}>M^2$ and
$M^2/20$.
For comparison, also the ${}^3\!S_1^{[8]}$-channel contributions to the latter
are shown.}
\end{figure*}

In direct $\gamma\gamma$ collisions, $J/\psi$ mesons with mass $M=2m_c$ and
finite transverse momentum $p_T$ are produced in association with either a
photon (see Fig.\ \ref{fig:1})
or a jet, where the final state photon in Fig.\ \ref{fig:1} has to be replaced
by a gluon. In the first case, the physical color-singlet state ($^3S_1^{[1]}$
in spectroscopic notation) is produced directly, whereas in the second case the
intermediate color-octet state ($^3S_1^{[8]}$) transforms into the physical
$J/\psi$ through soft gluon emission. Although the strong coupling of the hard
gluon enhances the second process, it is strongly suppressed by the relevant
color-octet OME, which is subleading in $v$. In NLO of
$\alpha_s$, an
unresolved dijet system with invariant mass $s_{jj}$, originating from two
gluons or a light quark-antiquark pair, allows for the production of
intermediate $^1S_0^{[8]}$, $^3S_1^{[8]}$, and $^3P_J^{[8]}$ states. At small
$p_T$, virtual loop corrections have to be included to cancel the unphysical
dependence on the cut-off $s_{jj}>M^2$, but at large $p_T$ these contributions
become unimportant. As can be observed clearly in Fig.\ \ref{fig:2}, the
cut-off dependence is reduced and the corrections from real particle emission
become large \cite{Klasen:2001mi}. At small $p_T$, one does, however, not
expect large radiative corrections, as has also been shown in a full
NLO calculation for direct color-singlet $J/\psi$
photoproduction \cite{Kramer:1994zi}.

\section{Comparison of $J/\psi$ production in complete $\gamma\gamma$ collisions with CERN LEP2 data}

$J/\psi$ production in $\gamma\gamma$ collisions proceeds not only through
direct interactions of photons with charm quarks, but also through resolved
processes, where one or two of the photons emit quarks and gluons which
then interact with the heavy quarks. In the small-$p_T$ range recently
studied by the DELPHI experiment at CERN LEP2 \cite{delphi}, $J/\psi$
production is dominated by the single-resolved process $\gamma g\to J/\psi g$,
where one initial state photon and the final state photon in Fig.\ \ref{fig:1}
have to be replaced with gluons. In addition, processes with non-abelian
coupling of the gluons and processes involving light quarks have to be taken
into account using the relevant color-octet OMEs. Furthermore, contributions
from $\chi_{cJ}$ and $\psi^\prime$ decays have to be included, whereas those
from $B$ meson decays are suppressed by the small branching fraction.

In our numerical analysis, we use $m_c=(1.5\pm0.1)$~GeV, $\alpha=1/137.036$,
and the LO formula for $\alpha_s^{(n_f)}(\mu)$ with $n_f=3$
active quark flavors.
As for the photon PDFs, we use the LO set from Gl\"uck, Reya, and Schienbein
(GRS) \cite{Gluck:1999ub}, which is the only available one that is implemented
in the fixed-flavor-number scheme, with $n_f=3$.
We choose the renormalization and factorization scales to be $\mu=\xi_\mu m_T$
and $\mu_f=\xi_f m_T$, respectively, where $m_T=\sqrt{M^2+p_T^2}$ is the
transverse mass of the $J/\psi$ meson, and independently vary the scale parameters
$\xi_\mu$ and $\xi_f$ between 1/2 and 2 about the default value 1.
As for the $J/\psi$, $\chi_{cJ}$, and $\psi^\prime$ OMEs, we adopt the set
determined in Ref.\ \cite{Braaten:1999qk} by fitting the Tevatron data
using the LO proton PDFs from Martin, Roberts, Stirling, and Thorne (MRST98LO)
\cite{Martin:1998sq} as our default and the one referring to the LO proton PDFs from
the CTEQ Collaboration (CTEQ5L) \cite{Lai:1999wy} for comparison (see Table~I
in
Ref.\ \cite{Braaten:1999qk}).
In the first (second) case, we employ $\Lambda_{\rm QCD}^{(3)}=204$~MeV
(224~MeV), which corresponds to $\Lambda_{\rm QCD}^{(4)}=174$~MeV \cite{Martin:1998sq}
(192~MeV \cite{Lai:1999wy}), so as to conform with the fit \cite{Braaten:1999qk}.
Incidentally, the GRS photon PDFs are also implemented with
$\Lambda_{\rm QCD}^{(3)}=204$~MeV \cite{Gluck:1999ub}.
In the cases $\psi=J/\psi,\psi^\prime$, the fit results for
$\langle{\cal O}^\psi[{}^1\!S_0^{[8]}]\rangle$ and
$\langle{\cal O}^\psi[{}^3\!P_0^{[8]}]\rangle$ are 
strongly correlated, and one is only sensitive to the linear combination
$
M_r^\psi=\langle{\cal O}^\psi[{}^1\!S_0^{[8]}]\rangle
+\frac{r}{m_c^2}
\langle{\cal O}^\psi[{}^3\!P_0^{[8]}]\rangle,
$
with an appropriate value of $r$.
Since the cross section is sensitive to a linear combination of
$\langle{\cal O}^\psi[{}^1\!S_0^{[8]}]\rangle$ and
$\langle{\cal O}^\psi[{}^3\!P_0^{[8]}]\rangle$ that differs from 
$M_r^\psi$, we write
$\langle{\cal O}^\psi[{}^1\!S_0^{[8]}]\rangle=\kappa
M_r^\psi$
and
$\langle{\cal O}^\psi[{}^3\!P_0^{[8]}]\rangle=(1-\kappa)
\left(m_c^2/r\right)M_r^\psi$ and vary $\kappa$ between 0 and 1 about the
default value 1/2.
The $J$-dependent OMEs
$\langle{\cal O}^\psi[{}^3\!P_J^{[8]}]\rangle$,
$\langle{\cal O}^{\chi_{cJ}}[{}^3\!P_J^{[1]}]\rangle$, 
and
$\langle{\cal O}^{\chi_{cJ}}[{}^3\!S_1^{[8]}]\rangle$
satisfy multiplicity relations,
which follow to leading order in $v$ from heavy-quark spin symmetry.
In order to estimate the theoretical uncertainties in our predictions, we
vary the unphysical parameters $\xi_\mu$, $\xi_f$, and $\kappa$ as indicated
above, take into account the experimental errors on $m_c$, the decay branching
fractions, and the default OMEs, and switch from our default OME set to the
CTEQ5L one, properly adjusting $\Lambda_{\rm QCD}^{(3)}$.
We then combine the individual shifts in quadrature, allowing for the upper
and lower half-errors to be different.

\begin{figure*}[htb]
\begin{center}
\epsfig{file=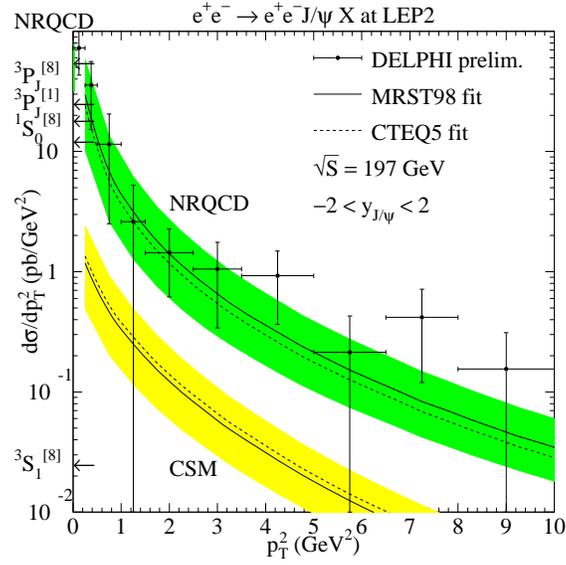,width=0.5\textwidth}
\end{center}
\caption{\label{fig:3}
The cross section $d\sigma/dp_T^2$ of $e^+e^-\to e^+e^-J/\psi+X$
measured by DELPHI \protect\cite{delphi} as a function of $p_T^2$ is compared
with the theoretical predictions of NRQCD and the CSM.
The solid and dashed lines represent the central predictions obtained with the
ME sets referring to the MRST98LO \protect\cite{Martin:1998sq} (default) and CTEQ5L
\protect\cite{Lai:1999wy} PDFs, respectively, while the shaded bands indicate the
theoretical uncertainties on the default predictions.
The arrows indicate the NRQCD prediction for $p_T=0$ and its 
${}^3\!P_J^{[1]}$, ${}^1\!S_0^{[8]}$, ${}^3\!S_1^{[8]}$, and ${}^3\!P_J^{[8]}$
components.}
\end{figure*}

In Fig.\ \ref{fig:3} we confront the $p_T^2$ distribution of
$e^+e^-\to e^+e^-J/\psi+X$ measured by DELPHI \cite{delphi} with our
NRQCD and CSM predictions.
The solid lines and shaded bands represent the central results, evaluated with
our default settings, and their uncertainties.
The DELPHI data clearly favors the NRQCD prediction, while it
significantly overshoots the CSM one \cite{Klasen:2001cu}.

\section{Conclusion}
In summary,
the production of $J/\psi$ mesons in $\gamma\gamma$ collisions allows for a
test of factorization in color-octet processes as predicted by NRQCD and
observed at the Tevatron. We have calculated the contributions
from direct and resolved photons to $J/\psi$ mesons produced directly or from
$\chi_{cJ}$ and $\psi^\prime$ decays. Our NRQCD predictions are nicely
confirmed by recent data from the DELPHI collaboration at CERN LEP2, whereas
the CSM prediction is clearly disfavored.

\section*{Acknowledgment}
The author thanks the organizers of the QCD Moriond 2003 conference for the
kind invitation and B.A.\ Kniehl, L.\ Mihaila, and M.\ Steinhauser for a
fruitful collaboration.

\end{document}